\documentclass[leqno,11pt]{amsart} 
\usepackage{amsfonts,epsfig,latexsym,amssymb,amsmath,graphics,color,float,rotating,lscape,hyperref} 
  
\newtheorem{lemma}{Lemma}[section] 
 
\newtheorem{theorem}[lemma]{Theorem} 
\newtheorem{proposition}[lemma]{Proposition} 
\newtheorem{definition}[lemma]{Definition} 
\newtheorem{corollary}[lemma]{Corollary} 
\newtheorem{example}[lemma]{Example} 
\newtheorem{exercise}[lemma]{Exercise} 
\newtheorem{remark}[lemma]{Remark} 
\newtheorem{fig}[lemma]{Figure} 
\newtheorem{tab}[lemma]{Table} 
  
\newcommand{\bth}{\begin{theorem}} 
        \newcommand{\ethe}{\end{theorem}} 
\newcommand{\bre}{\begin{remark}\em } 
        \newcommand{\ere}{\end{remark}} 
\newcommand{\ble}{\begin{lemma}} 
        \newcommand{\ele}{\end{lemma}} 
\newcommand{\bde}{\begin{definition}} 
        \newcommand{\ede}{\end{definition}} 
\newcommand{\bco}{\begin{corollary}} 
        \newcommand{\eco}{\end{corollary}} 
  
\newcommand{\bpr}{\begin{proposition}} 
        \newcommand{\epr}{\end{proposition}} 
  
\newcommand{\bexer}{\begin{exercise}} 
        \newcommand{\eexer}{\end{exercise}} 
  
\newcommand{\bexam}{\begin{example}} 
        \newcommand{\eexam}{\end{example}} 
  
\newcommand{\bfi}{\begin{fig}} 
        \newcommand{\efi}{\end{fig}} 
  
\newcommand{\btab}{\begin{tab}} 
        \newcommand{\etab}{\end{tab}} 

\def\B_e{B_{\eta}(e)}

\definecolor{darkblue}{rgb}{.1, 0.1,.8} 
\definecolor{darkgreen}{rgb}{0,0.8,0.2} 
\definecolor{darkred}{rgb}{.8, .1,.1}

\newcommand{\beao}{\begin{eqnarray*}} 
        \newcommand{\eeao}{\end{eqnarray*}\noindent} 
  
\newcommand{\beam}{\begin{eqnarray}} 
\newcommand{\eeam}{\end{eqnarray}\noindent} 
  
\newcommand{\beqq}{\begin{equation}} 
\newcommand{\eeqq}{\end{equation}\noindent} 
  
\newcommand{\bce}{\begin{center}} 
        \newcommand{\ece}{\end{center}} 
  
\newcommand{\barr}{\begin{array}} 
        \newcommand{\earr}{\end{array}}

\newcommand{\vague}{\stackrel{\lower0.2ex\hbox{$\scriptscriptstyle 
                        \it{v} $}}{\rightarrow}} 
\newcommand{\dist}{\stackrel{\lower0.2ex\hbox{$\scriptscriptstyle 
                        \it{d} $}}{\rightarrow}} 
\newcommand{\weak}{\stackrel{\lower0.2ex\hbox{$\scriptscriptstyle 
                        \it{w} $}}{\rightarrow}} 
\newcommand{\what}{\stackrel{\lower0.2ex\hbox{$\scriptscriptstyle 
                        \it{\hat{w}} $}}{\rightarrow}} 
  
\newcommand{\bdis}{\begin{displaymath}} 
\newcommand{\edis}{\end{displaymath}\noindent}

\begin{document}

	\title[Dynamic investment portfolio optimization]{Dynamic investment portfolio optimization using a Multivariate Merton Model with Correlated Jump Risk}

	\author[B. Afhami]{Bahareh Afhami}
	\author[M. Rezapour]{Mohsen Rezapour}
		\author[M. Madadi]{Mohsen Madadi}

		\author[V. Maroufy]{Vahed Maroufy}
	\address[B. Afhami]{Department of Statistics, Faculty of Mathematics and Computer, Shahid Bahonar University of Kerman, Kerman, Iran}\email{baharehafhami@math.uk.ac.ir}
	\address[M. Rezapour]{Corresponding author:
		Department of Biostatistics \& Data Science, School of Public Health, The University of Texas Health Science Center at Houston (UTHealth),
		Houston, Texas }\email{mohsenrzp@gmail.com}
	\address[M. Madadi]{Department of Statistics, Faculty of Mathematics and Computer, Shahid Bahonar University of Kerman, Kerman, Iran}\email{madadi@uk.ac.ir}
	\address[V. Maroufy]{
		Department of Biostatistics \& Data Science, School of Public Health, The University of Texas Health Science Center at Houston (UTHealth),
		Houston, Texas }\email{Vahed.Maroufy@uth.tmc.edu}
	

\begin{abstract}
	In this paper, we are concerned with the optimization of a dynamic investment portfolio  when the   securities which follow a multivariate Merton model with dependent jumps are periodically invested and proceed by approximating the Condition-Value-at-Risk (CVaR) by comonotonic bounds and maximize the expected terminal wealth. Numerical studies as well as applications of our results to real datasets are also provided.\\
Keywords:{ 
Risk analysis, Conditional tail expectation, Merton Model, Geometric Brownian motion, Comonotonicity.
}\\
		2010 Mathematics Subject Classification: Primary C630; \ \ Secondary C580;\ C650
		
	\end{abstract}
    
        \thanks{} 
        \maketitle 
  
\section{Introduction} 
One of the most important problems in investment strategies is choosing the portfolio that ideally leads to the  maximum return but has minimum risk. This problem was first proposed by Markowitz in 1950 \cite{Markowitz:1952}, who used a quadratic programming model for solving this optimization problem. Since then, the problem has been studied by several researchers using different criteria. For example, 
Cura \cite{Cura:2009} used the particle swarm optimization method, while Xu et al. \cite{Xu:2017} assumed a Black-Scholes market with stochastic drift and used the fractional Kelly-strategy developed by MacLean et al. \cite{MacLean:2006}. Furthermore, Afhami et al. \cite{Afhami} used comonotonic approximation to solve the problem when the securities followed a multivariate Merton model with dependent jumps and the portfolio was constant. In this paper, we use the fractional Kelly-strategy to optimize wealth in a jump-diffusion Merton model with a dynamic control asset allocation (DCAA). The solution leads to a stochastic differential equation (SDE), which was solved for the constant mix strategy in Afhami et al. \cite{Afhami} and can be generalized to the DCAA strategy to obtain the terminal wealth of a periodic investment problem in a DCAA strategy framework. The optimization problem is solved based on a lower bound of the risk measure of terminal wealth using comonotonic theory. 

The rest of the paper is organized as follows. Section \ref{Preliminaries} provides the definitions and existing results used throughout the paper. 
The main results are presented in Section \ref{Main results}, we first obtain the solution of the SDE corresponding to the terminal wealth process of a portfolio following a jump-diffusion model using the DCAA strategy, and then calculate the lower comonotonic bound for CVaR of the terminal wealth in order to solve the portfolio optimization problem.
 In Section \ref{Data Example}, we implement our optimization method on data containing the daily prices of the stocks of Zoom Video Communication Inc. (ZM), and Tesla Inc. (TSLA). These data are readily available from the $S\&P\, 500$ market index.

\section{{Preliminaries}} \label{Preliminaries}
The Black-Scholes model is a mathematical model used for options pricing and estimating the variation of financial investments over time. The model assumes that price of assets follows a geometric Brownian motion with constant drift and volatility, and is widely used in the literature. For 
Al-Zhour et al. \cite{Zhour:2019} investigated  pricing using a nonlinear volatility model. 
Jankova \cite{Jankova:2018} focused on the methods of derivative contract pricing, and derived the basic differential equation of the Black-Scholes model for option contract pricing. He also referred to the significant drawbacks and limitations of other option pricing models which are based on unrealistic assumptions. 
 Merton \cite{Merton:1973} showed that the Black-Scholes model is generally not suitable for real market stock pricing, and introduced the jump-diffusion ``Merton" model \cite{Merton:1976} which is commonly used in the literature and has abundant applications in the analysis of financial data. For example, 
 St\"{u}binger et al. \cite{Stubinger:2017} developed a pairs trading framework based on a mean-reverting jump-diffusion model and applied it to the $S\&P\, 500$ oil companies.
  Li et al. \cite{Li:2018} introduced a dynamic model for a risky asset which is governed by its compensation process, a Brownian motion and a stationary compound Poisson process. 
  Maruddani et al. \cite{Maruddani:2019} applied jump-diffusion processes to derive bond parameters, equity and default probability, when the asset prices have extreme values. 
  In this paper, we consider the return of a portfolio including of one risk-free asset together with several risky assets which follow the jump-diffusion Merton model and optimize it using the DCAA strategy.

$VaR$ and $CVaR$ are the most common risk measures used in an optimization problem for maximizing expected terminal wealth. 
\begin{definition}
Let $X$ be a random variable with cdf $F_X$ the risk measures $VaR,\, CVaR$ and $CLVaR,$ for the set of real numbers $\mathbb{R}$ and $p\in (0,1),$ are defined as 
\begin{eqnarray*} 
VaR_{p}(X)&:=&\inf \{ x\in \mathbb{R} \mid F_{X}(x) \geq p \},\\
CVaR_{p}(X)&:=&\mathbb{E} \left( X \mid X >VaR_{p}(X) \right), \\
CLVaR_{p}(X)&:=&\mathbb{E} \left( X \mid X < VaR_{p}(X) \right).  
\end{eqnarray*} 
where $F_{X}(x)=Pr(X \leq x) $ and by convention, $\inf \{\emptyset \} = + \infty.$ 
\end{definition}
It is obvious that
$$CVaR_{1-p}(X)=-CLVaR_{p}(-X). $$

The following definition characterizes comonotonicity (see  Theorem 3 in  Dhaene et al. \cite{Dhaene:2002a}).

\bde
The random vector ${\mathbf{X}}=(X_1, X_2, \ldots, X_n)$ is comonotonic if and only if one of the following equivalent conditions holds:
\begin{itemize}
\item[$(1)$] ${\mathbf{X}}$ has a comonotonic support;
\item[$(2)$] For all ${\mathbf{x}}=(x_1, x_2, \ldots, x_n),$ we have 
\begin{eqnarray*}
F_{{\mathbf{X}}}({\mathbf{x}})=\min\{F_{X_1}(x_1),F_{X_2}(x_2),\ldots,F_{X_n}(x_n) \};
\end{eqnarray*}
\item[$(3)$] For $U\sim Uniform(0,1),$ we have 
\begin{eqnarray*}
\hspace{-1.7cm} {\mathbf{X}}\overset{d}{=}(F_{X_1}^{-1}(U),F_{X_2}^{-1}(U),\ldots, F_{X_n}^{-1}(U) );
\end{eqnarray*}
\item[$(4)$]There exist a random variable $Z$ and non-decreasing functions $f_i,\, (i=1,2,\ldots,n),$ such that
\begin{eqnarray*}
\hspace{-2.7cm} {\mathbf{X}}\overset{d}{=}(f_1(Z),f_2(Z),\ldots, f_n(Z)).
\end{eqnarray*} 
\end{itemize}
\ede

The following theorem presents additivity property of risk measures for sums of comonotonic risks (Dhaene et al. \cite{Dhaene:2006}).
\bth
For random vector $(X_1,X_2,\ldots,X_n)$ that is comonotonic we have 
\begin{eqnarray*} 
&&VaR_p(\sum\limits_{j=1}^n X_j)=\sum\limits_{j=1}^n VaR_p(X_j),\\
&&CVaR_p(\sum\limits_{j=1}^n X_j)=\sum\limits_{j=1}^n CVaR_p(X_j), 
\end{eqnarray*} 
if all marginal distributions $F_k,\, (k=1,2,\ldots,n)$ are continuous.
\ethe

A random variable $X$ is said to precede $Y$ in the convex order sense $(X \leq_{cx} Y)$, if $\mathbb{E}(X) =\mathbb{E}(Y)$ and $\mathbb{E}[\max(X-d , 0)] \leq \mathbb{E}[\max(Y-d , 0)]$ for all real $ d.$ 
With respect to convex order, Kaas et al. \cite{Kaas:2000} obtains bounds for sum of random variables.
\bth \label{Convex657}
 For any random vector 
$(X_1,\ldots,X_n)$ and any random variable $\Lambda$, 
\begin{eqnarray*} 
\sum\limits_{j=1}^{n}\mathbb{E}(X_{j}\mid \Lambda) \leq_{cx} \sum\limits_{j=1}^{n}X_{j} \leq_{cx} \sum\limits_{j=1}^{n}F_{X_{j}}^{-1}(U). 
\end{eqnarray*} 
\ethe

\section{Main results}\label{Main results} 
In this section, we consider $(m + 1)$ securities consisting of one risk-free and $m$ risky assets that are traded continuously. We also assume that the decision maker invests a fraction $x_j(l-1), \,(j=1,2,\ldots,m)$ of his/her wealth in the $j$-th risky asset at the beginning of the $l$-th period and $1-\sum\limits_{j=1}^m x_j(l-1)$ in the risk-free asset. In what follows, Theorem \ref{theorem31} gives the solution of the corresponding SDE (Eq. \ref{stochastic equation}) and Theorem \ref{jkhu776} characterizes the solution to the optimization problem. All  notations are provided in Table \ref{jnkg8} for simplicity.
\begin{small}
\begin{table}
\center \caption{Notations used throughout this paper} \label{jnkg8}
\begin{tabular}{ c| c  }
    \noalign{\smallskip} \hline
    $Symbol$ & $Definition$    \\  \hline 
    $\mathcal{P}_j(t)$ & Price\, of\, the\, $j^{th}$\, risky\, asset\, at\, time\, $t$   \\    \rule{0pt}{6mm}
   $\mu_j$ & Drift\, of\, geometric\, Brownian\, motion   \\    \rule{0pt}{6mm}
    $\sigma_j$ & Volatility\, of\, the\, asset\, $\mathcal{P}_j,\,\, j=1,2,\ldots,m $  \\     \rule{0pt}{6mm}   
 ${\mathbf{\mu}}$ & $(r+\mu_1-\lambda h_{1,0}-\lambda_1 h_{1,1},\ldots,r+\mu_m-\lambda   h_{m,0}-\lambda_m h_{m,1})^\top$ \\  \rule{0pt}{6mm}
    ${\mathbf {x}}(l-1)$ & $(x_1(l-1),\ldots,x_m(l-1))^\top$    \\    \rule{0pt}{6mm}
     $r$ & Interest\, rate\, of \, risk-free \,asset  \\   \rule{0pt}{6mm}
     $ {\mathbf{A}}$ & ${\mathbf{\mu}}-r {\mathbf{1}}$    \\     \rule{0pt}{6mm}
    $\mu({\mathbf{x}}(l-1))$ & ${\mathbf{A}}^\top\, {\mathbf{x}}(l-1)+r$  \\   \rule{0pt}{6mm}
    $ \mathfrak{B'_j}$ &  m-dimensional\, standard\, Brownian\, motions,\,\,   \\  
    $ $ & $j=1,2,\ldots,m$ \\  \rule{0pt}{6mm}
    $\rho_{i,j}$ & $\frac{1}{t}Cov(\mathfrak{B'_i}(t),\mathfrak{B'_j}(t+s))\,,\,\, t,s\geq 0,\,   i,j=1,\ldots,m$  \\    \rule{0pt}{6mm}
    ${\mathbf{\Sigma}}_{m\times m}$ & ${\left(\sigma_{i}\,\sigma_{j}\,\rho_{i,j}\right)}_{i,j}\,,\, i,j\in \{1,2,\ldots,m\}$  \\   \rule{0pt}{6mm}
    $\sigma^2({\mathbf{x}}(l-1))$ & $({\mathbf{x}}(l-1))^\top\,{\mathbf{\Sigma}}\,\, {\mathbf{x}}(l-1)$  \\  \rule{0pt}{6mm}
    $\mathfrak{B}_{{\mathbf{x}}(l-1)}(t)$ & $(\sigma^2({\mathbf{x}}(l-1)))^{-1}\,  \sum\limits_{j=1}^{m}  x_j(l-1)\, \sigma_j\,\mathfrak{B'_j}(t)$  \\   \rule{0pt}{6mm}
    $\mathcal{N}(t)$ & Poisson\, processes\, with\, intensity\, rates\, $\lambda$  \\   \rule{0pt}{6mm}
    $\mathcal{N}_j(t)$ & Poisson\, processes\, with\, intensity\, rates\, $\lambda_j$ \\   \rule{0pt}{6mm}
    $ \mathcal{Z}_{k,j,0}$ &  The\, jump\, magnitude\, of\, the\, $k$-th \, common\, jump  \\ 
    $ $ & for\, the\,asset\, $\mathcal{P}_j\, in\, (0,t],\,\, k=1,2,\ldots,\mathcal{N}(t) $\\   \rule{0pt}{6mm}
    $ \mathcal{Z}_{k,j,1}$ & The\, jump\, magnitude\, of\, the\, $k$-th\, individual\, jump\,  \\ 
    $ $ &  of\, asset\, $\mathcal{P}_j,\, in\, (0,t],\,\, k=1,2,\ldots,\mathcal{N}_j(t)$ \\   \rule{0pt}{6mm}   
    $h_{j,\ell}$ & $E[e^{\mathcal{Z}_{\iota(t),j,\ell}}]-1,\,\, j=1,2,\ldots,m ,\, \ell=0,1$   \\    \rule{0pt}{5mm}
    $\alpha_{l^\star}$ & The\, periodic\, endowments\, at\, predetermined  \\ 
    $ $ & $ points,\,\, l^\star=0,1,\ldots,\tau-1$  \\   \rule{0pt}{6mm}
    $\mu_{\mathcal{Z}_{1,j,\ell}}$ & $\mathbb{E}(\mathcal{Z}_{k,j,\ell})$ \\   \rule{0pt}{6mm}
     $\sigma^2_{\mathcal{Z}_{1,j,\ell}}$& $Var(\mathcal{Z}_{k,j,\ell})$ \\   \rule{0pt}{6mm}
    $r_{n}$ & $corr(V_n, \Lambda)$ \\   \rule{0pt}{6mm}
    $ w_{l-1}$ & The\, amount\, of \,wealth\, at\, the\, beginning\, of\, the\, period\, $l$\\     \hline   
\end{tabular}
\end{table}
\end{small}

In the constant mix strategy the initial proportions of assets will remain fixed until the end of the terminal horizon.  
However, when using the DCAA strategy, the optimal portfolio strategy is obtained based on the current wealth to design a policy that remains fixed until the end of the terminal horizon \cite{Maruddani:2019}. If at the start of the next period the amount of wealth changes,  the strategy is updated. 

Assume that $\mathcal{P}_0(t),$ the price of the risk-free asset at time $t,$ satisfies 
\begin{eqnarray}\label{equation1jgvkj} 
\frac{d\mathcal{P}_0(t)}{\mathcal{P}_0(t- )}=r\, dt, \hspace{0.5cm} r>0,
\end{eqnarray} 
and the dynamics of risky assets prices for $j=1,2,\ldots,m,$ follow 
\begin{eqnarray} \label{equation1} \nonumber 
\frac{d \mathcal{P}_j(t)}{\mathcal{P}_j(t-)}&=&(r+\mu_j -\lambda\, h_{j,0}-\lambda_j\, h_{j,1})\, dt+\sigma_j\, d\mathfrak{B'_j}(t)+(e^{\mathcal{Z}_{\iota(t-),j,0}}-1)\, d\mathcal{N}(t)\\ 
&&+(e^{\mathcal{Z}_{\iota(t-),j,1}}-1)\, d\mathcal{N}_j(t). 
\end{eqnarray} 

Discounted changes in assets price are divided into two groups: individual changes corresponding to $\mathcal{N}_j(t)$ and common changes corresponding to $\mathcal{N}(t).$ 
We assume that for all $j,\, k,$ and $\ell=0,1$ the random variables $\mathcal{Z}_{k,j,\ell}$ are independent of $\mathfrak{B'_j}(t)$ and for fixed $j=1,2,\ldots,m,$ the random variables $\mathcal{Z}_{k,j,\ell}$ are $i.i.d$ for all $k$ and $\ell.$ The variables $\mathcal{Z}_{k,j,\ell}$ and $\mathcal{Z}_{k,j',\ell}$ are also independent for $j\neq j',$ but they may not be identically distributed. We assume that $\mathcal{Z}_{k,j,\ell},\, (j=1,\ldots,m,\, \ell=0,1)$ are continuous random variables.
    
Then, according to (\ref{equation1}) and (\ref{equation1jgvkj}), the wealth process $W(t)$ within the $l$-th period is 
\begin{eqnarray}\label{stochastic equation}\nonumber 
\frac{d \mathcal{P}(t)}{\mathcal{P}(t-)}&=&\sum\limits_{j=1}^{m}x_j(l-1)\, \frac{d \mathcal{P}_j(t)}{\mathcal{P}_j(t-)}+(1-\sum\limits_{j=1}^{m}x_j(l-1)) \frac{d \mathcal{P}_0(t)}{\mathcal{P}_0(t-)}\\ \nonumber 
&=&\mu\left({\mathbf{x}}(l-1)\right)\,dt+\sigma^2({\mathbf{x}}(l-1))\,d \mathfrak{B}_{{\mathbf{x}}(l-1)}(t) 
+\sum\limits_{j=1}^{m} x_j(l-1)\, (e^{\mathcal{Z}_{\iota(t-),j,0}}-1)\,d\mathcal{N}(t)\\
&&+\sum\limits_{j=1}^{m} x_j(l-1)\,(e^{\mathcal{Z}_{\iota(t-),j,1}}-1)\,d\mathcal{N}_j(t). 
\end{eqnarray} 

With a simple modification of Proposition 2.1 in Afhami et al. \cite{Afhami} we can obtain the solution of the Equation 
(\ref{stochastic equation}), and because of the similarity of proof, we do not provide the proof here.

\bth\label{theorem31}
If the random processes $\mathcal{N},\mathcal{N}_{j_1}$ and $\mathcal{N}_{j_2}$ do not have common jumps for $j_1\neq j_2=1,2,\ldots,m,$ then the solution of (\ref{stochastic equation}) is as bellow
 \begin{small}
\begin{eqnarray} \label{hhk77}  \nonumber
\mathcal{P}(t)&=& e^{\left(\mu({\mathbf{x}}(l-1))-
\frac{\sigma^2({\mathbf{x}}(l-1))} 
{2}\right)\,t+\sigma^2({\mathbf{x}}(l-1))\, \mathfrak{B}_{{\mathbf{x}}(l-1)}(t)+\sum\limits_{j=1}^{m}  \sum\limits_{k=1}^{\mathcal{N}(t)}\mathcal{Z}^\star_{k,j,0}({{\mathbf{x}}(l-1)})}\\
&&\times e^{\sum\limits_{j=1}^{m} \sum\limits_{k=1}^{\mathcal{N}_j(t)}\mathcal{Z}^\star_{k,j,1}({\mathbf{x}}(l-1))}. 
\end{eqnarray}
\end{small} 
where $e^{\mathcal{Z}^\star_{k,j,\ell}({\mathbf{x}}(l-1))}-1=x_j(l-1) (e^{\mathcal{Z}_{k,j,\ell}}-1),\, \ell=0,1,\, j=1,\ldots,m.$ 
\ethe

An investment of a unit amount of wealth at time $t-1$ will grow to amount $e^{\mathcal{Y}_{t}({{\mathbf{x}}(l-1)})}$ at time $t,$ where 
\begin{small}
\begin{eqnarray}\label{ypil}\nonumber
&&\mathcal{Y}_{t}({\mathbf{x}}(l-1))=\mu({\mathbf{x}}(l-1))-
\frac{
\sigma^2({\mathbf{x}}(l-1))
}{2}+\sum\limits_{j=1}^{m}  \sum\limits_{k=N(t-1)+1}^{\mathcal{N}(t)}\mathcal{Z}^\star_{k,j,0}({\mathbf{x}}(l-1)) \\ 
&&+\sum\limits_{j=1}^{m} \sum\limits_{k=\mathcal{N}_j(t-1)+1}^{\mathcal{N}_j(t)}\mathcal{Z}^\star_{k,j,1}({\mathbf{x}}(l-1))
+\sigma^2({\mathbf{x}}(l-1))\, \left(\mathfrak{B}_{{\mathbf{x}}(l-1)}(t)-\mathfrak{B}_{{\mathbf{x}}(l-1)}(t-1)\right). 
\end{eqnarray} 
\end{small}

In Afhami et al. \cite{Afhami}, the terminal wealth was calculated from \eqref{hhk77} assuming constant proportions and an optimal strategy was obtained to maximize the expectation of terminal wealth when the comonotonic approximation of $CLVaR$, was controlled. We now consider portfolio optimization under DCAA strategy.

So the wealth $\mathcal{W}_l$ at the end of the $l$-th period  satisfy the recursion equation 
\begin{eqnarray*}  
\mathcal{W}_l=\mathcal{W}_{l-1}\,e^{\mathcal{Y}_{l}({\mathbf{x}}(l-1))}+\alpha_l, \hspace{.75cm} l=1,2,\ldots, \tau, 
\end{eqnarray*} 
with $\mathcal{W}_0=\alpha_0.$ Hence the terminal wealth can be written as 
\begin{eqnarray}\label{kjjjjjjj}
\mathcal{W}_\tau({\mathbf{x}}(l-1))=w_{l-1} \, e^{\sum\limits_{k=l}^{\tau}\mathcal{Y}_{k}({\mathbf{x}}(l-1))}
+\sum\limits_{j=l}^{\tau-1} \alpha_j \, e^{\sum\limits_{k=j+1}^{\tau}\mathcal{Y}_{k}({\mathbf{x}}(l-1))}.
\end{eqnarray} 

Due to the limitation of borrowing from risky assets, we impose the upper bound $c_0$ to $\mu({\mathbf{x}}(l-1))\,,$  so the portfolio optimization problem is given by: 
\begin{eqnarray} \label{optimization11}
&& \max\limits_{{\mathbf{x}}(l-1)}\, \mathbb{E}\left(\mathcal{W}_\tau({\mathbf{x}}(l-1)) \right),\\ \nonumber 
&& CLVaR_{p}\left(\mathcal{W}_\tau({\mathbf{x}}(l-1)) \right)\geq K,\\ \nonumber
&&\mu({\mathbf{x}}(l-1)) \leq c_0.\nonumber 
\end{eqnarray} 

Since we cannot obtain the exact distribution of $\mathcal{W}_\tau,$ instead we approximate its comonotonic lower bound.
\bth \label{uuio9}
Let $\Lambda=\sigma^2({\mathbf{x}}(l-1))\,
\sum\limits_{k=l-1}^{\tau-1} \alpha_k\,
 \left(\mathfrak{B}_{{\mathbf{x}}(l-1)}(\tau)-\mathfrak{B}_{{\mathbf{x}}(l-1)}(k) \right),$
then the comonotonic lower bound is 
\begin{eqnarray} \label{jjo8} \nonumber
\mathcal{W}_{\tau}^{L}({\mathbf{x}}(l-1))&=& w_{l-1}\, 
e^{c_{1,l-1}+c_{2,l-1}({\mathbf{x}}(l-1))+
c_{3,l-1} \, \sqrt{\sigma^2({\mathbf{x}}(l-1))}\,\,  \frac{\Lambda}{\sigma_\Lambda}}\\ 
&&+\sum\limits_{i=l}^{\tau-1} \alpha_i \, e^{c_{1,i}+c_{2,i}({\mathbf{x}}(l-1))
+ c_{3,i}\, \sqrt{\sigma^2({\mathbf{x}}(l-1))} \,\,  
\frac{\Lambda}{\sigma_\Lambda}}\, ,
\end{eqnarray}
where $c_{1,n},\, c_{2,n},$ and $ c_{3,n}$ are defined in (\ref{kkk1}), (\ref{kkk2}) and (\ref{kkk3}).
\ethe

Calculating risk measures of (\ref{jjo8}) is difficult, to come up with this issue, we use the first-order Taylor expansion of the exponential function. Its result is given in follow Lemma.
\begin{lemma} \label{lemma44}
We can approximate $\mathcal{W}_{\tau}^{L}({\mathbf{x}}(l-1))$ using the first-order Taylor expansion as
\begin{eqnarray}\label{uuu10}
\mathcal{W}_{\tau}^{'L}({\mathbf{x}}(l-1))=c_4+c_5({\mathbf{x}}(l-1))+c_6\, 
\, \sqrt{\sigma^2({\mathbf{x}}(l-1))}\, \frac{\Lambda}{\sigma_\Lambda},
\end{eqnarray}
where $c_4,\, c_5({\mathbf{x}}(l-1))$ and $c_6$ are given in (\ref{hh78}), (\ref{hh79}) and (\ref{hh80}).
\end{lemma}

One can easily shown that 
 ${\mathbb{E}}(\mathcal{W}_{\tau}^{'L})=c_4+c_5({\mathbf{x}}(l-1)),$ and its risk measure is 
\begin{eqnarray*}
CVaR_{1-p}(-\mathcal{W}_{\tau}^{'L})=-c_4-c_5({\mathbf{x}}(l-1))+c_7\,
 \sqrt{\sigma^2({\mathbf{x}}(l-1))},
\end{eqnarray*}
where $c_7=c_6\, \frac{1}{p\, \sqrt{2\, \pi}}\, e^{-\frac{1}{2}\, (\Phi^{-1}(p) )^2}.$

So portfolio optimization problem reduces to
\begin{eqnarray}\label{hjk895}
&&\max\limits_{{\mathbf{x}}(l-1)} \left(c_4+c_5({\mathbf{x}}(l-1)) \right),\\ \nonumber
&&-c_4-c_5({\mathbf{x}}(l-1))+c_7\, \sqrt{\sigma^2({\mathbf{x}}(l-1))}\leq -K, \\ \nonumber
&& r+{\mathbf{A}}^\top {\mathbf{x}}(l-1)\leq c_0,
\end{eqnarray}
and its solution is given in the following result.

\bth \label{jkhu776}
The solution of problem (2.9) for $p<0.5$ is of the form ${\mathbf{x}}(l-1)=q(l-1) \, {\mathbf{x}}^\star,$ where ${\mathbf{x}}^\star={\mathbf{\Sigma}}^{-1}\, {\mathbf{A}},$ 
$q(l-1)=\min\{q_1,\, q_2,\, q_3 \},$ and
\begin{eqnarray}
q_1=\frac{-B_2-\sqrt{B_2^2-4\, B_1\, B_3} }{2\, B_1}, \hspace{.5cm}
q_2=\frac{c_8}{2\, c_9}, \hspace{.5cm}
q_3=\frac{c_0-r}{{\mathbf{A}}^\top {\mathbf{\Sigma}}^{-1}\,\, {\mathbf{A}} },
\end{eqnarray}
with
\begin{eqnarray*}
B_1=-c_9\,\left( {\mathbf{A}}^\top {\mathbf{\Sigma}}^{-1}\,\, {\mathbf{A}}\right),\hspace{.5cm}
B_2=\frac{c_8}{-c_9}\, B_1-c_7\, \sqrt{\frac{B_1}{-c_9}},\hspace{.5cm}
B_3=c_4-K,
\end{eqnarray*}
and
\begin{eqnarray*}
&&c_8=w_{l-1}\, (\tau-l+1)+\sum\limits_{i=l}^{\tau-1} \alpha_i\, (\tau-i),\\
&&c_9=\frac{1}{2}\, w_{l-1}\, (\tau-l+1)\, r^2_{l-1}+\frac{1}{2}\, \sum\limits_{i=l}^{\tau-1} \alpha_i\,r^2_{i}.
\end{eqnarray*}
\ethe

\section{Data Example} \label{Data Example}
In Example \ref{tttt444}, we compute the optimal portfolio for a DCAA strategy for two stocks listed in the $S\&P\, 500$ stock market index.

\begin{figure}[ht]
\centerline{\includegraphics[height=6 cm, width=8 cm]{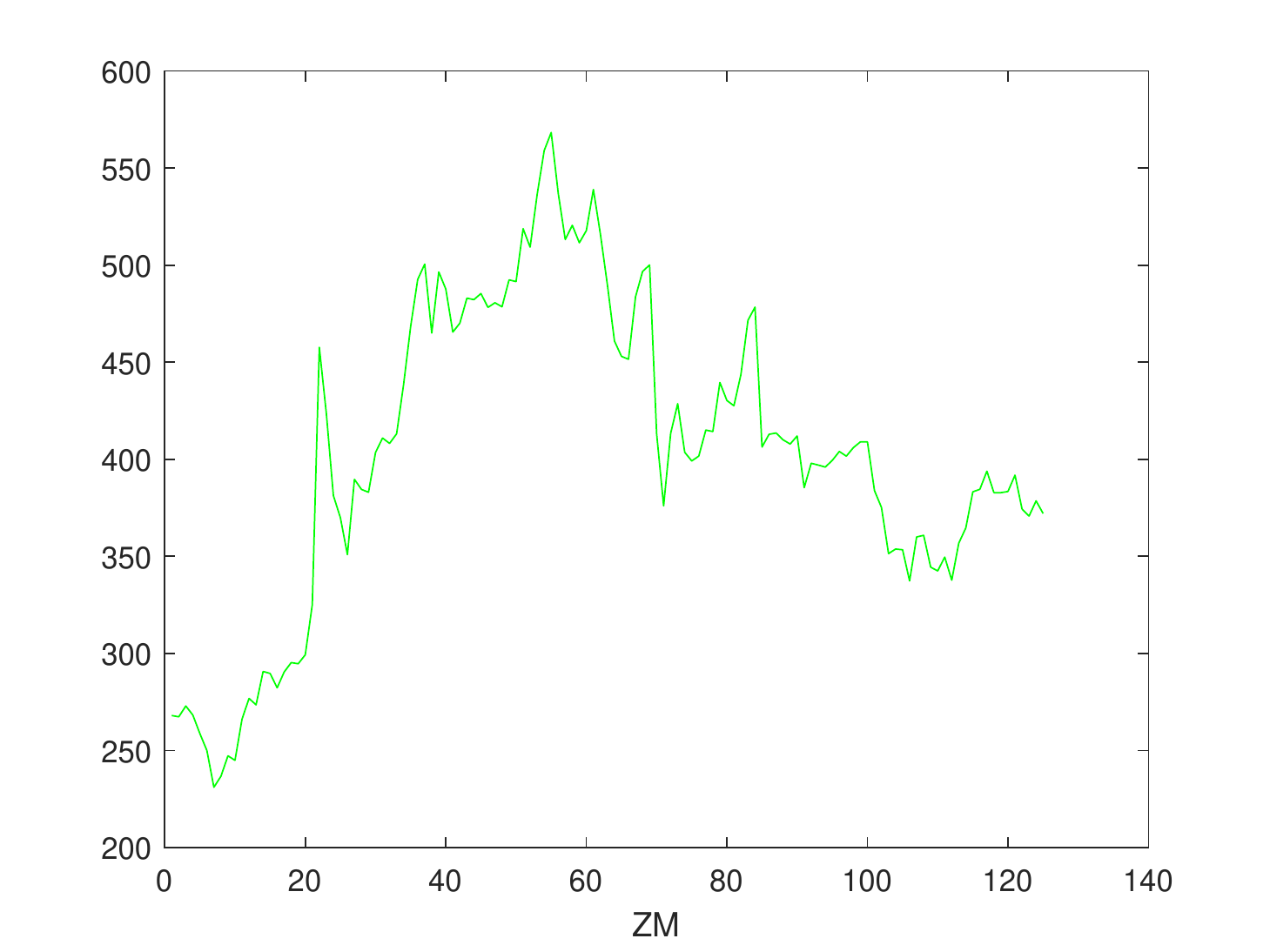}
\includegraphics[height=6 cm, width=8 cm]{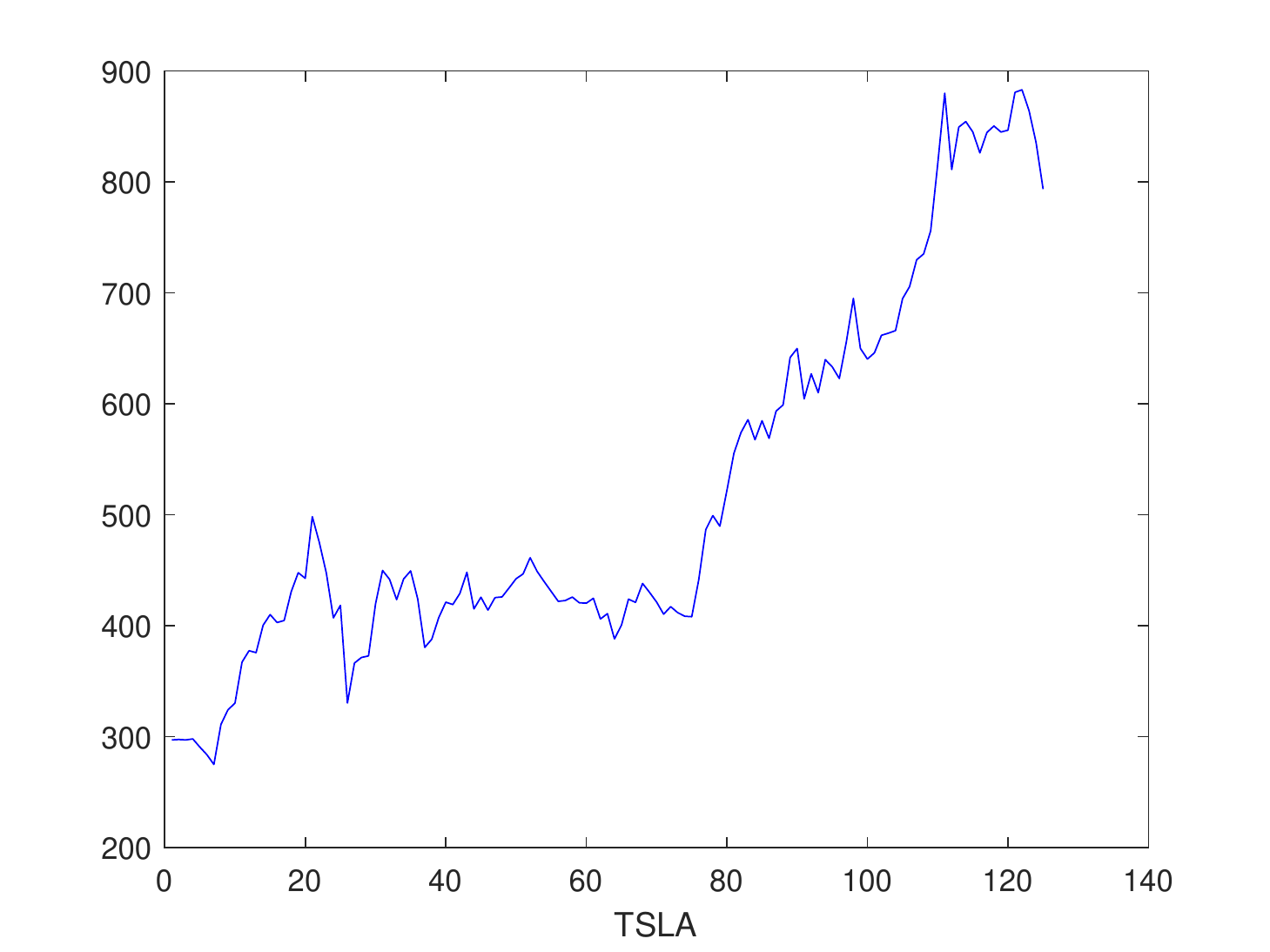}}
\caption{\label{e2plot} Diagram of daily prices of assets.}
\end{figure}

\bexam\label{examreal} \label{tttt444}
In this example, we consider the daily prices of  ZM and TSLA for time horizon $\tau=6$ months from August of $2020$ to January of $2021.$ These prices are plotted in Figure \ref{e2plot}. 

For simplification we set $\alpha_0=\alpha_1=\ldots=\alpha_5=1,$ and assume that the interest rate and the risk budget are $r = 0.03$ and $p = 0.05,$ respectively. 
In  Equation (\ref{hjk895}) we set $K=k^\star \sum\limits_{i=1}^{\tau} e^{\frac{(\tau-i+1)\, r}{\tau}} $ where $k^\star$ is the stop loss rate. From Theorem \ref{jkhu776} we obtain the optimal proportions $x_1(l)$ and $x_2(l)$ respectively of the stocks ZM and TSLA  at the beginning of the $l$-th period i.e. at time $l^\star =l-1$ and write ${\mathbf{x}}(l^\star)=(x_1(l^\star),x_2(l^\star)),\, (l^\star=0,1,\ldots,5).$

Let $ {\mathbf{S}}_b (l)=(S_{b,1}(l),S_{b,2}(l))$ and ${\mathbf{S}}_e (l)=(S_{e,1}(l),S_{e,2}(l))$ be vectors respectively containing the prices of ZM and TSLA at the beginning and end of the $l$-th period, respectively. 
\begin{figure}[ht]
\centerline{\includegraphics[width=11cm]{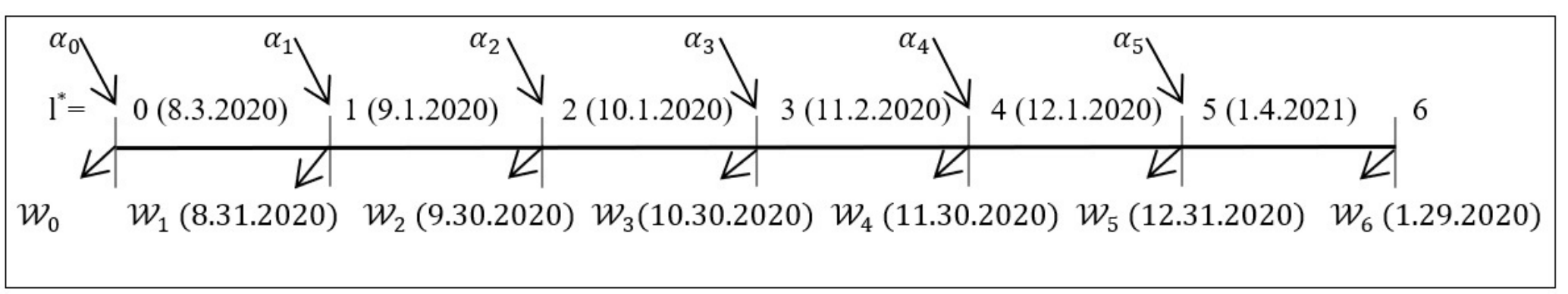}}
\caption{\label{e5678ot1} Illustration of the algorithm for six periods.}
\end{figure}

Consider ${\mathbf{f}}_l=(\frac{\mathcal{W}_l \, x_1(l)}{S_{b,1}(l)}, \frac{\mathcal{W}_l \, x_2(l)}{S_{b,2}(l)})$ where $\frac{\mathcal{W}_l \, x_1(l)}{S_{b,1}(l)}$ and $\frac{\mathcal{W}_l \, x_2(l)}{S_{b,2}(l)}$ are respectively the stock volumes of ZM and TSLA which are bought at the beginning of the $l$-th period. We compute terminal wealth ($\mathcal{W}_6$) as follows: 

{\bf{First step}}: At time $l^\star=0\, (or\, l=1),$ with $\mathcal{W}_0=1$ we obtain ${\mathbf{x}}(0)$ using Theorem \ref{jkhu776}. 

{\bf{Second step}}: At time $l^\star=1 \, (or\, l=2),$ we calculate $\mathcal{W}_1={\mathbf{f}}_0\,  {\mathbf{S}}_e (0)^\top+\alpha_1,$ and again using Theorem \ref{jkhu776} obtain ${\mathbf{x}}(1).$ 

{\bf{Third step}}: We repeat second step for $l^\star=2,3,\ldots,5 \, (or\, l=3,4,\ldots,6),$ with $\mathcal{W}_{l^\star}={\mathbf{f}}_{l^\star-1}\,  {\mathbf{S}}_e (l^\star-1)^\top+\alpha_{l^\star},$ and using Theorem \ref{jkhu776} to obtain ${\mathbf{x}}(l^\star).$

This algorithm is illustrated in Figure \ref{e5678ot1}. 

The results for different values of $k^\star$ are given in Table \ref{jkh3}. This table illustrates that wealth and return of portfolio decreases as $k^\star$ increases.
\begin{table}
 \center \caption{Optimal wealth for the first six periods.} \label{jkh3}
 \begin{tabular}{ c| c c c c c c c}
    \noalign{\smallskip} \hline
    $k^\star$ & $\mathcal{W}_1$  & $\mathcal{W}_2$ & $\mathcal{W}_3$ & $\mathcal{W}_4$ & $\mathcal{W}_5$ & $\mathcal{W}_6$ & $Return$ \\  \hline  
    $0.50$ &  $2.4640$ &$3.3657$ &$4.0509$ & $6.0675$ & $7.2707$ & $7.7254$ & $26.5194 \%$ \\
    $0.85$ &  $2.3898$  &$3.2944$ &$3.9863$ & $5.9867$ & $7.1871$ & $7.6366$ & $25.0651 \%$ \\  
    $0.90$ &  $2.3118$  &$3.0252$ &$3.7422$ & $5.6813$ & $6.8715$ & $7.3012$ & $19.5722 \%$ \\
    $0.95$ &  $2.2326$  &$2.7155$ &$3.4355$ & $5.2977$ & $6.4751$ & $6.8801$ & $12.6758 \%$ \\ \hline                               
 \end{tabular}
 \end{table}
\eexam

{\bf{Acknowledgement:}} The authors are grateful to Professor Mahbanoo Tata for editing this paper as well as for her valuable suggestions. 
\vskip.3in

\newpage 

{\small

}

\newpage

\appendix 
\section{Appendix} 

{\bf{Proof of Theorem \ref{uuio9}: }} 

\begin{proof}
From (\ref{kjjjjjjj}) and (\ref{ypil}) it can be easily shown
\begin{eqnarray*}
\mathcal{W}_\tau({\mathbf{x}}(l-1))=w_{l-1}\, e^{V_{l-1}+S_{l-1}}+\sum\limits_{i=l}^{\tau-1} \alpha_i\, e^{V_{i}+S_{i}},
\end{eqnarray*}  
where for $n=l-1,l,\ldots,\tau-1,$
\begin{eqnarray*}
&& V_{n}=\sigma^2({\mathbf{x}}(l-1))  
 \left(\mathfrak{B}_{{\mathbf{x}}(l-1)}(\tau)-\mathfrak{B}_{{\mathbf{x}}(l-1)}(n) \right),\\ 
&&S_{n}=(\tau-n)\left(\mu({\mathbf{x}}(l-1))-\frac{\sigma^2({\mathbf{x}}(l-1))}{2} \right)
+\sum\limits_{j=1}^{m} \sum\limits_{k=N(n)+1}^{N(\tau)} \mathcal{Z}^\star_{k,j,0}({\mathbf{x}}(l-1))\\
&&\hspace{1cm} +\sum\limits_{j=1}^{m} \sum\limits_{k=\mathcal{N}_j (n)+1}^{\mathcal{N}_j (\tau)} \mathcal{Z}^\star_{k,j,1}({\mathbf{x}}(l-1)), \\
\end{eqnarray*}
By using Theorem \ref{Convex657}, we obtain lower bound for $\mathcal{W}_\tau({\mathbf{x}}(l-1)).$ First we set 

$\Lambda=\sigma^2({\mathbf{x}}(l-1))\, \sum\limits_{k=l-1}^{\tau-1} \alpha_k\,
 \left(\mathfrak{B}_{{\mathbf{x}}(l-1)}(\tau)-\mathfrak{B}_{{\mathbf{x}}(l-1)}(k) \right),$
so
\begin{eqnarray*}
&&\sigma^2_{\Lambda}=\sigma^2({\mathbf{x}}(l-1))\,
\sum\limits_{k,w=l-1}^{\tau-1}\alpha_k\, \alpha_w\, \min\{\tau-k,\tau-w \},\\
&&\sigma^2_{V_n}=(\tau-n)\, \sigma^2({\mathbf{x}}(l-1)),
\end{eqnarray*}
moreover
\begin{eqnarray*}
r_{n}=Corr(V_{n},\Lambda )=\frac{\sum\limits_{k=l-1}^{\tau-1} \alpha_k\, \min\{\tau-n,\tau-k\} }
{\sqrt{(\tau-n)\,\sum\limits_{k,w=l-1}^{\tau-1} \alpha_k\, \alpha_w\, \min\{\tau-k,\tau-w \} } }.
\end{eqnarray*}
Therefore the lower bound is
\begin{eqnarray}\label{uuu1}
\mathcal{W}_{\tau}^{L}({\mathbf{x}}(l-1))&=&{\mathbb{E}}\left(w_{l-1}\, e^{V_{l-1}+S_{l-1}} 
+\sum\limits_{i=l}^{\tau-1} \alpha_i\, e^{V_{i}+S_{i}} \mid \Lambda \right)\\  \nonumber
&&=w_{l-1}\, {\mathbb{E}}(e^{V_{l-1}}\mid \Lambda )\, {\mathbb{E}}(e^{S_{l-1}})
+\sum\limits_{i=l}^{\tau-1} \alpha_i\, {\mathbb{E}}(e^{V_{i}}\mid \Lambda)\, {\mathbb{E}}(e^{S_{i}}).
\end{eqnarray}
Since $V_{n}\mid \Lambda=\lambda $ has standard normal distribution with mean and variance, respectively 
$r_{n}\, \sigma_{V_n}\frac{\lambda}{\sigma_\Lambda},$ and 
$\sigma_{V_n}^2\, (1-r_{n}^2).$  
We have
\begin{eqnarray*}\label{uuu3} \nonumber
{\mathbb{E}}(e^{V_{n}}\mid \Lambda )=e^{r_{n}\,  
\sqrt{(\tau-n)\, \sigma^2({\mathbf{x}}(l-1))}\, \frac{\Lambda}{\sigma_\Lambda}\,
+\frac{1}{2} (\tau-n)\, \sigma^2({\mathbf{x}}(l-1))(1-r^2_{n})},
\end{eqnarray*}
and
\begin{eqnarray*}\label{uuu5} \nonumber
{\mathbb{E}}(e^{S_{n}})&=& e^{(\tau-n)  \mu({\mathbf{x}}(l-1))-\frac{1}{2} (\tau-n)\,
\sigma^2({\mathbf{x}}(l-1)) 
+\lambda (\tau-n) \left(\prod\limits_{j=1}^{m} M_{\mathcal{Z}_{1,j,0}^\star({\mathbf{x}}(l-1))}(1)-1 \right)}\\ 
&&\times e^{(\tau-n) \sum\limits_{j=1}^{m} \lambda_j \left(M_{\mathcal{Z}_{1,j,1}^\star({\mathbf{x}}(l-1))}(1)-1 \right)}.
\end{eqnarray*}
Thus
\begin{eqnarray*} 
\mathcal{W}_{\tau}^{L}({\mathbf{x}}(l-1))&=& w_{l-1}\, 
e^{c_{1,l-1}+c_{2,l-1}({\mathbf{x}}(l-1))+
c_{3,l-1} \, \sqrt{\sigma^2({\mathbf{x}}(l-1))}\,\,  \frac{\Lambda}{\sigma_\Lambda}}\\ 
&&+\sum\limits_{i=l}^{\tau-1} \alpha_i \, e^{c_{1,i}+c_{2,i}({\mathbf{x}}(l-1))
+ c_{3,i}\, \sqrt{\sigma^2({\mathbf{x}}(l-1))} \,\,  
\frac{\Lambda}{\sigma_\Lambda}},
\end{eqnarray*}
where
\begin{eqnarray}\label{kkk1}\nonumber
\hspace{-2cm} c_{1,n}&=&(\tau-n) r +\lambda (\tau-n) 
\left(\prod\limits_{j=1}^{m} M_{\mathcal{Z}_{1,j,0}^\star({\mathbf{x}}(l-1))}(1)-1 \right)\\
\hspace{-2cm} &&+(\tau-n) \sum\limits_{j=1}^{m} \lambda_j 
\left(M_{\mathcal{Z}_{1,j,1}^\star({\mathbf{x}}(l-1))}(1)-1 \right),
\end{eqnarray}
\begin{eqnarray}\label{kkk2}
\, c_{2,n}({\mathbf{x}}(l-1))=(\tau-n)\left( {\mathbf{A}}^\top {\mathbf{x}}(l-1) \right)
-\frac{1}{2}\, (\tau-n)\, r_{n}^2\, \sigma^2({\mathbf{x}}(l-1)),
\end{eqnarray}
\begin{eqnarray}\label{kkk3}
\hspace{-7.8cm} c_{3,n}= \sqrt{(\tau-n)}\,\, r_{n},
\end{eqnarray}
which completes the proof of the result.
\end{proof}

{\bf{Proof of Lemma \ref{lemma44}: }} 
Here, we use the first-order Taylor expansion of the exponential function in (\ref{jjo8}). Therefore,  
\begin{small}
\begin{eqnarray*}
&&\mathcal{W}_{\tau}^{'L}({\mathbf{x}}(l-1))=w_{l-1}\, 
\left(1+c_{1,l-1}+c_{2,l-1}({\mathbf{x}}(l-1))
+ c_{3,l-1} \, \sqrt{\sigma^2({\mathbf{x}}(l-1))}\, \frac{\Lambda}{\sigma_\Lambda} \, \right)\\
&&+\sum\limits_{i=l}^{\tau-1} \alpha_i\, \left(1+c_{1,i}+c_{2,i}({\mathbf{x}}(l-1))
+c_{3,i}\, 
+\frac{\Lambda}{\sigma_\Lambda}\, \sqrt{\sigma^2({\mathbf{x}}(l-1))}  
\right).
\end{eqnarray*}
\end{small}
By straightforward calculation, we have
\begin{eqnarray*}
\mathcal{W}_{\tau}^{'L}({\mathbf{x}}(l-1))=c_4+c_5({\mathbf{x}}(l-1))+c_6\, 
\, \sqrt{\sigma^2({\mathbf{x}}(l-1))}\, \frac{\Lambda}{\sigma_\Lambda},
\end{eqnarray*}
where
\begin{eqnarray} \label{hh78}
\hspace{-3cm} c_4=w_{l-1}\, (1+c_{1,l-1})+\sum\limits_{i=l}^{\tau-1} \alpha_i\, (1+c_{1,i}),
\end{eqnarray}
\begin{eqnarray} \label{hh79}
c_5({\mathbf{x}}(l-1))=w_{l-1}\, c_{2,l-1}({\mathbf{x}}(l-1))+\sum\limits_{i=l}^{\tau-1} \alpha_i\, c_{2,i}({\mathbf{x}}(l-1)),
\end{eqnarray}
and
\begin{eqnarray} \label{hh80}
\hspace{-5cm} c_6=w_{l-1}\, c_{3,l-1}+\sum\limits_{i=l}^{\tau-1} \alpha_i\, c_{3,i}. 
\end{eqnarray}

This completes the proof of results.
\end{document}